\documentclass[reprint,amsmath,amssymb,aps,prl]{revtex4-1}

\usepackage{graphicx}%
\usepackage{multirow}%
\usepackage{amsmath,amssymb,amsfonts}%
\usepackage{amsthm}%
\usepackage{mathrsfs}%
\usepackage[title]{appendix}%
\usepackage{xcolor}%
\usepackage{notes2bib}

\begin{document}

\preprint{APS/123-QED}

\title{Extremely weak sub-kelvin electron-phonon coupling in InAs On Insulator}

\author{Sebastiano Battisti}
\email{sebastiano.battisti@sns.it}
\affiliation{NEST, Istituto Nanoscienze-CNR and Scuola Normale Superiore, I-56127 Pisa, Italy}

\author{Giorgio De Simoni}
\affiliation{NEST, Istituto Nanoscienze-CNR and Scuola Normale Superiore, I-56127 Pisa, Italy}

\author{Alessandro Braggio}
\affiliation{NEST, Istituto Nanoscienze-CNR and Scuola Normale Superiore, I-56127 Pisa, Italy}

\author{Alessandro Paghi}
\affiliation{NEST, Istituto Nanoscienze-CNR and Scuola Normale Superiore, I-56127 Pisa, Italy}

\author{Lucia Sorba}
\affiliation{NEST, Istituto Nanoscienze-CNR and Scuola Normale Superiore, I-56127 Pisa, Italy}

\author{Francesco Giazotto}
\affiliation{NEST, Istituto Nanoscienze-CNR and Scuola Normale Superiore, I-56127 Pisa, Italy}

\begin{abstract}
   We are proposing, as an ideal candidate for caloritronic devices operating at subKelvin temperatures, a hybrid superconductor-semiconductor platform named InAs on insulator (InAsOI). This heterostructure is made by doped InAs grown on an insulating buffer of InAlAs on a GaAs substrate. Caloritronic devices aim to heat or cool electrons out of equilibrium with respect to the phonon degree of freedom. However, their performances are usually limited by the strength of the electron-phonon (e-ph) coupling and the associated power loss. Our work discusses the advantages of the InAsOI platform, which are based on the significantly low e-ph coupling measured compared to all-metallic state-of-the-art caloritronic devices. Our structure demonstrates values of the e-ph coupling constant up to two orders of magnitude smaller than typical values in metallic structures.
\end{abstract}

\maketitle

In the past decade, significant attention has been paid to studying heat management in superconducting micro-structures, also known as coherent caloritronics \cite{fornieri2017}. The objective is to propose the equivalent of coherent electronics in superconductive circuits \cite{giazotto2006,hwang2020,linder2016}. Different types of caloritronic devices have been proposed as fundamental components for potential fully coherent, heat-driven superconductive circuits and memory elements \cite{wang2008,ordonez-miranda2019,kubytskyi2014,germanese2022}, thermal rectifiers \cite{fornieri2015,li2004}, transistors \cite{paolucci2017}, thermal logic gates \cite{wang2007,paolucci2018}, and thermal switches \cite{liu2024}. 
These devices operate at very low temperatures below one Kelvin, where the interaction between electrons and phonons weakens. This means that the temperature of the electrons can be very different from that of the phonons (i.e., the lattice). This characteristic is essential for caloritronic devices. The ability to significantly alter the electronic temperature compared to the phonon temperature enables all caloritronic devices to have a wider operational range and overall better performance.\\
Past research has explored hot electron effects and electron-phonon (e-ph) interaction in semiconductors \cite{bauer1972,price1982,muhonen2011}. It is known that semiconductors have a weaker e-ph interaction compared to metals. However, a significant drawback for coherent caloritronics and Josephson physics applications is the non-Ohmic barrier they typically create with superconductive metals. In this study, we focus on investigating a recently proposed superconductor-semiconductor hybrid platform based on an InAs heterostructure grown on an insulating buffer of InAlAs, known as InAs On Insulator (InAsOI) \footnote{F. Giazotto, G. De Simoni, L. Sorba, A. Paghi, O. Arif, and C. Puglia, ''Josephson Junction and superconductor field effect transistor'', Filling number: 102024000008983 (19/04/2024)} \cite{paghi2024}, which, as we will show, has an extremely weak e-ph interaction. 
This can be demonstrated by the minimal power needed to significantly raise the electronic temperature out of equilibrium compared to the temperature of the phonon bath. Additionally, our platform forms a clean, Ohmic contact with metals, resulting in a robust proximity effect \cite{paghi2024}. This enables the incorporation of superconductive Josephson physics within the caloritronic framework with relative ease.\\ 
To measure the electron-phonon interaction, we use a superconductor-normal metal-superconductor (S-N-S) Josephson junction (JJ) as an electronic thermometer to monitor the equilibrium electron temperature. Simultaneously, we apply a constant heating power to the structure. By repeating this measurement at different temperatures of the phonon bath, we can determine the electron-phonon interaction using a power law fit of $T^n$ as described in Wellstood and co-workers \cite{Wellstood1994}. 

\begin{figure}[t!]
  \includegraphics[width=1\columnwidth, trim= 0cm 1cm 0cm 0cm]{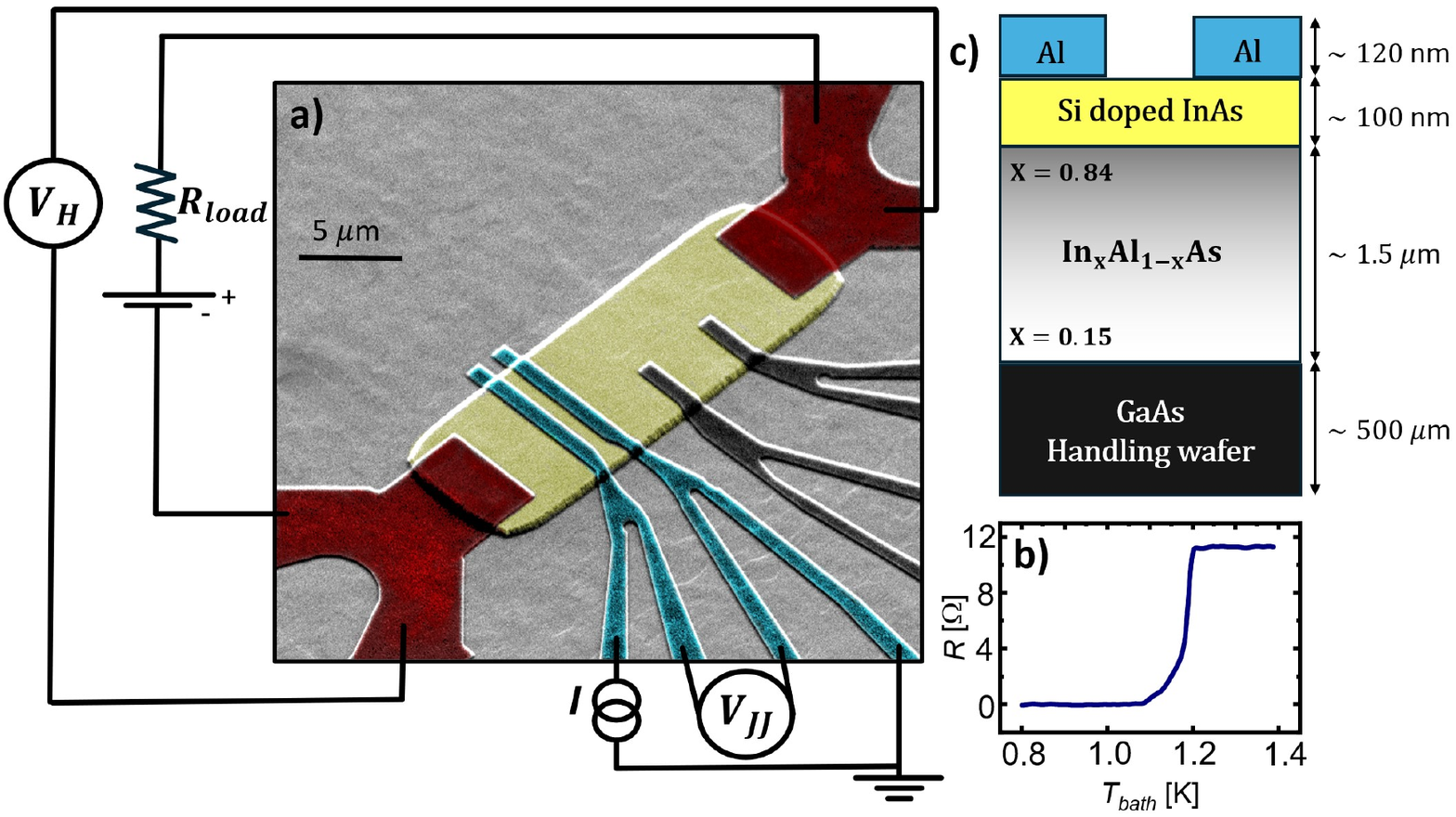}
    \caption{
    \textbf{(a)}: 
    Scanning electron micrograph of a representative sample from this study showing the wiring used for measurements. The red contacts represent the Joule heaters, the blue ones are the leads of the Josephson junction separated by approximately 550 nm, and the yellow area indicates the InAs mesa.  \textbf{(b)}: 
    Resistance as a function of the temperature of the Josephson junction in device 1. \textbf{(c)}: Diagram of the InAsOI heterostructure used, with the thickness of each layer shown on the right.
    }
  \label{Fig1}
\end{figure}

The devices were fabricated starting from the InAsOI substrate shown in Fig.\ref{Fig1}(c). A $100$-nm-thick $n^{++}$ Si-doped InAs epitaxial layer with three-dimensional charge carrier concentration (extracted via room-temperature Hall measurements) of $n_{3D}=(2.53\pm0.02)\times 10^{18}$ cm$^{-3}$ is grown on top of an InAlAs metamorphic buffer layer, having a graded In concentration to reduce the strain between the GaAs substrate and the InAs layer\cite{arif2024,benali2022}.\\
The samples were first cleaned and then prepared by the definition of electron beam lithography (EBL) Ti/Au ($10$/$40$ nm) markers for subsequent re-aligned lithographies. Then InAs mesas were defined through negative EBL and chemical etching with a standard \text{III-V} etching solution H$_2$O/C$_6$H$_8$O$_7$/H$_2$O$_2$/H$_3$PO$_4$ : $21/5/0.5/0.25$ vol (C$_6$H$_8$O$_7$  1M, H$_2$O$_2$ 30\% in mass and  
$H_3PO_4$ 85\% in mass, etching rate $\sim 100$ nm/s). After the definition of the mesas, the InAs has to pass through the surface passivation and Al evaporation process. This process is done by leaving the sample in sulfur supersaturated (NH$_4$)$_2$S$_x$ solution ($0.29$ M of (NH$_4$)$_2$S and $0.3$ M of S in DIW) at $45^{\circ}$C for $60$ s, which removes the native oxide on the InAs surface and replaces the oxygen atoms with sulfur ones until the sample is inserted in the electron-beam evaporator, where the S atoms are desorbed from the surface in the ultra-high vacuum (UHV). 
Consequently, the evaporation of $120$ nm of Al all over the sample creates a clean contact with the InAs, resulting in a robust superconducting proximity effect. 
Then, the Al contacts on the InAs mesa are defined by negative EBL and selective chemical etching of the Al (with Transene Al etchant type D). 
The result of the fabrication process is shown in Fig.\ref{Fig1}(a), where the Joule heaters (in red) and the JJ leads (in blue) are visible.\\
The InAs mesa (in yellow) is $\sim 20$ $\mu$m long and $\sim 6$ $\mu$m wide. The distance between the Joule heaters (in red) is $\sim 17$ $\mu$m, and the separation between the leads of the JJ is $\sim 550$ nm (in blue). 
The grey contacts were used to attempt another type of thermometry technique based on the zero bias anomaly at the contact between a superconductor and a disordered semiconductor \cite{giazotto2001,giazotto2001b}. However, this thermometer was overly sensitive to even minor magnetic field fluctuations, making this approach unstable and impractical. As a result, these contacts are not used in the results presented in this work.

\begin{figure}[t!]
  \includegraphics[width=1\columnwidth, trim= 0cm 0cm 0cm 0cm]{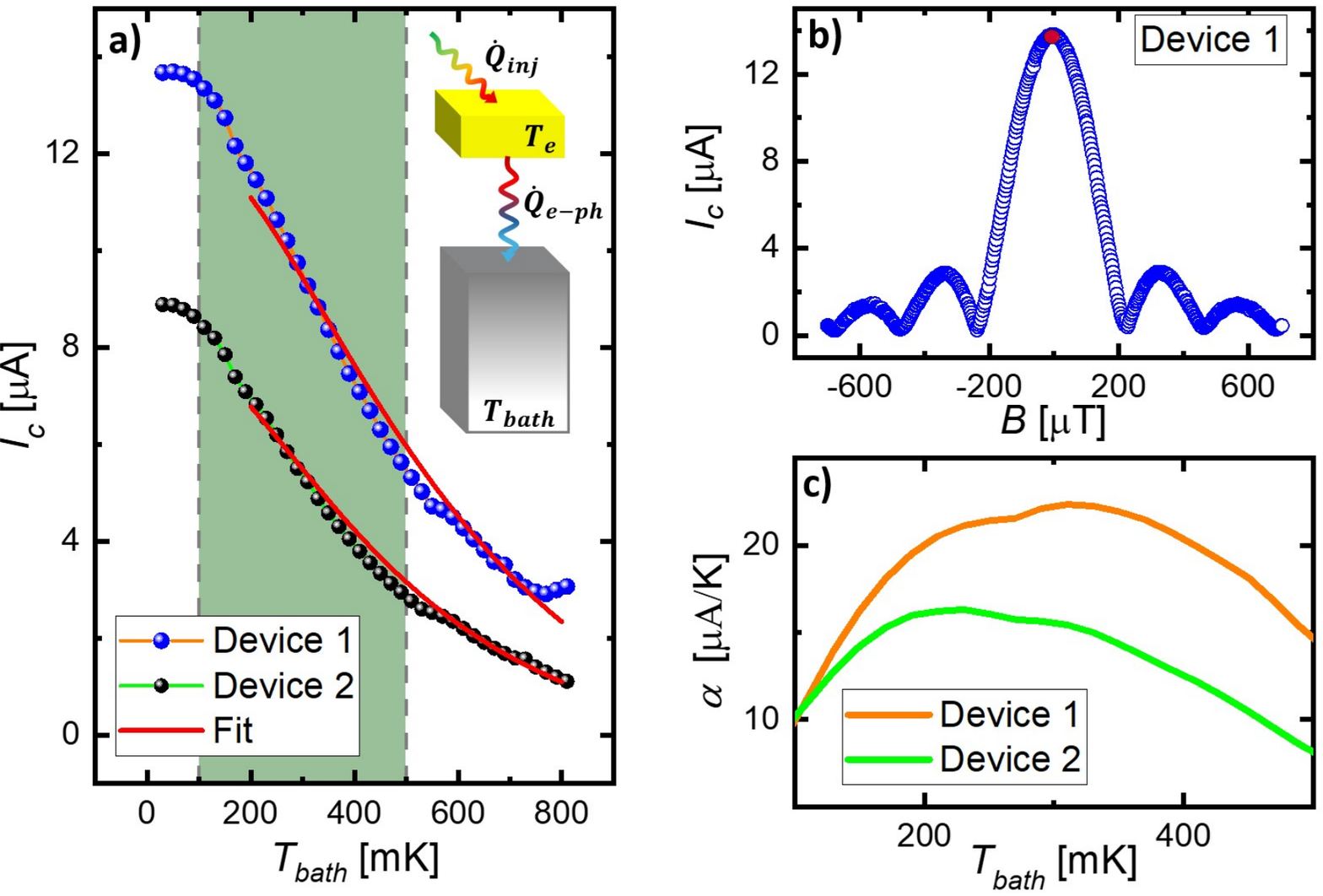}
    \caption{
    \textbf{(a)}: Critical current ($I_c$) of the Josephson junction as a function of bath temperature $T_{bath}$ for both devices. The red lines best fit Eq.\ref{eqJJ} for both devices. The green-shaded area highlights the region used for the thermometer calibration. 
    The inset shows a scheme of the predominant heat currents flowing through the InAsOI system: $\dot Q_{inj}$ is the injected power from the Joule heaters, whereas $\dot Q_{e-ph}$ represents the power dissipated to the substrate lattice phonons residing at the temperature of the cryostat.
    \textbf{(b)}: $I_c$ as a function of the out-of-plane magnetic field (Fraunhofer-like pattern) of device 1 taken at $T_{bath}=30$mK. 
    The red dot in the panel indicates the critical current value used for the temperature calibration curves of the thermometer in panel \textbf{(a)}.
    \textbf{(c)}: Temperature responsivity $\alpha =|\partial I_C/\partial T_{bath}|$ (absolute value of the first derivative of the curves in \textbf{(a)}) of the Josephson thermometers, as a function of $T_{bath}$ for both devices.}
  \label{Fig2}
\end{figure}

The devices were wired as shown in Fig.\ref{Fig1}(a) and cooled down in a cryogen-free dilution refrigerator to reach the superconducting temperature for Al. Figure \ref{Fig1}(b) shows the resistance of the JJ ($R$) as a function of the bath temperature ($T_{bath}$). It can be seen that the critical temperature of the JJ is $T_c\simeq 1.2$ K, which is approximately equal to the Al one $T_c^{Al}=1.196$ K \cite{cochran1958,paghi2024}. This originates from a very good proximitization stemming from the InAs Fermi level pinning above the conduction band when interfaced with a metal \cite{bhargava1997,melitz2010} and from the high doping of InAs, which makes it a degenerate semiconductor.\\
Figure \ref{Fig2}(a) shows the measured critical current $I_c$ as a function of $T_{bath}$ for the two samples considered in this study. These measurements are taken with the heaters off (so we assume that $T_{bath}=T_e$), and while $T_{bath}$ is changed, the critical current of the JJ is measured.  These represent the cornerstone characteristics of this work since they provide the calibration curves of the thermometers used to probe the InAs electronic temperature $T_e$ locally. To gain some insight into the effective length of the junctions, the curves in Fig.\ref{Fig2}(a) were fitted using the model for the critical current of a long diffusive JJ in the high-temperature regime, so considering only the experimental data that satisfy the condition $k_B T \gtrsim 5 E_{Th} $ \cite{dubos2001}:
\begin{equation}
    I_c(T)=\frac{64 \pi k_B T}{e R_N} \sum_{n=0}^{\infty} \frac{\sqrt{\frac{2\omega_n}{E_{Th}}} \Delta^2(T) \exp[-\sqrt{\frac{2\omega_n}{E_{Th}}}]}{[\omega_n + \Omega_n + \sqrt{2(\Omega_n^2 + \omega_n\Omega_n)}]^2},
    \label{eqJJ}
\end{equation}
where $E_{Th}$ is the junction Thouless energy left as the only fitting parameter, $k_B$ the Boltzmann constant, $R_N$ the normal-state resistance of the JJ, $\Delta(T)=1.764k_B T_c \tanh[{1.74\sqrt{(T_c/T)-1}}]$ is the temperature-dependent superconducting energy gap, $\omega_n=(2n+1)\pi k_B T$, and $\Omega_n=\sqrt{\omega_n^2 + \Delta^2(T)}$ are the Matsubara frequencies. The result of the fit is $E_{th_1}=4.7\cdot10^{-24}$J and $E_{th_2}=3.5\cdot 10^{-24}$J for device 1 and 2, respectively. 
On the other hand, $E_{Th}=\hbar D/L^2$ \cite{edwards1972}, where $L$ is the length of the JJ, $D=l_e v_F/3$ is the 3-dimensional diffusion coefficient, $l_e$ is the electron mean free path, and $v_F$ is the Fermi velocity. So one can find the effective lengths of the JJs via the fitted $E_{th_{1,2}}$, obtaining $L_1\simeq 1.43$ $\mu$m and $L_2\simeq1.71$ $\mu$m for devices 1 and 2, respectively. The discrepancy with the nominal length of the JJs, which is $\sim 550$ nm, has to be ascribed to the diffusive behavior of the InAsOI platform. Indeed, the length extracted by the fits might be taken as an effective length over which the correlated electrons forming Cooper pairs can travel from one lead to another. Moreover, it must be noted that the critical current values and fitting curves quality for device 2 and device 1 are quite different. This can be ascribed to the non-perfect equality of the JJs due to the small differences that may arise during fabrication. In particular, the passivation step and the contact definition by EBL may not produce identical outcomes, causing the devices to have slightly different properties. \\
The optimal operating region for the thermometer is the green-shaded area in Fig.\ref{Fig2}(a), ranging from $100$ mK to $500$ mK. This region is optimal due to the quasi-linear slope of the curve, allowing for a reliable numerical conversion between $I_c$ and $T_e$.
In other regions, the curve flattens, which makes it less suitable to define the conversion from $I_c$ to $T_e$.\\
Figure \ref{Fig2}(b) shows the Fraunhofer-like interference pattern of one of the JJs, which is constructed by measuring the $I_c$ of the JJ vs the out-of-plane magnetic field ($B$). 
This pattern demonstrates the coherence of the Josephson effect \cite{baronepaternò,suominen2017} and the capability to adjust the critical supercurrent of the JJ using an external magnetic field. 
The red spot in Fig.\ref{Fig2}(b) indicates the optimal magnetic field working point, around $B\sim 0$ T, for the calibration curves shown in Fig.\ref{Fig2}(a). 
This specific magnetic field value  was selected to achieve a higher $I_c$, resulting in a more significant response from the Josephson thermometer.\\
Figure \ref{Fig2}(c) shows the computed thermometer responsivity $\alpha=|\partial I_C/\partial T_{bath}|$ as a function of $T_{bath}$. This parameter represents the slope of the curves in Fig. \ref{Fig2}(a), measuring how sensitive the thermometer is to temperature variations. 
We can evaluate the Noise Equivalent Temperature ($NET$) (i.e., temperature sensitivity) of our thermometers as $NET=\sqrt{S_I}/\alpha$ \cite{gasparinetti2015} where the noise spectral density of the detected switching current $I_C$ can be evaluated as $S_I=(\delta I_C)^2/Bw$ with $\delta I_C\simeq30$ nA is the experimental uncertainty on the switching current and $Bw\simeq300$ Hz is the bandwidth of our setup. Using this estimation, one finds $NET\simeq 100-200$ $\mu$K/$\sqrt{\mathrm{Hz}}$ throughout the temperature range studied. This NET is comparable with state-of-the-art devices \cite{gasparinetti2015,giazotto2006}, making this platform appealing as a building block for calorimetric measurements and single photon detection \cite{virtanen2018,kokkoniemi2020,kokkoniemi2019,solinas2018}.\\
\begin{figure*}[t!]
  \includegraphics[width=1.7\columnwidth, trim= 2.7cm 0cm 2.2cm 3cm]{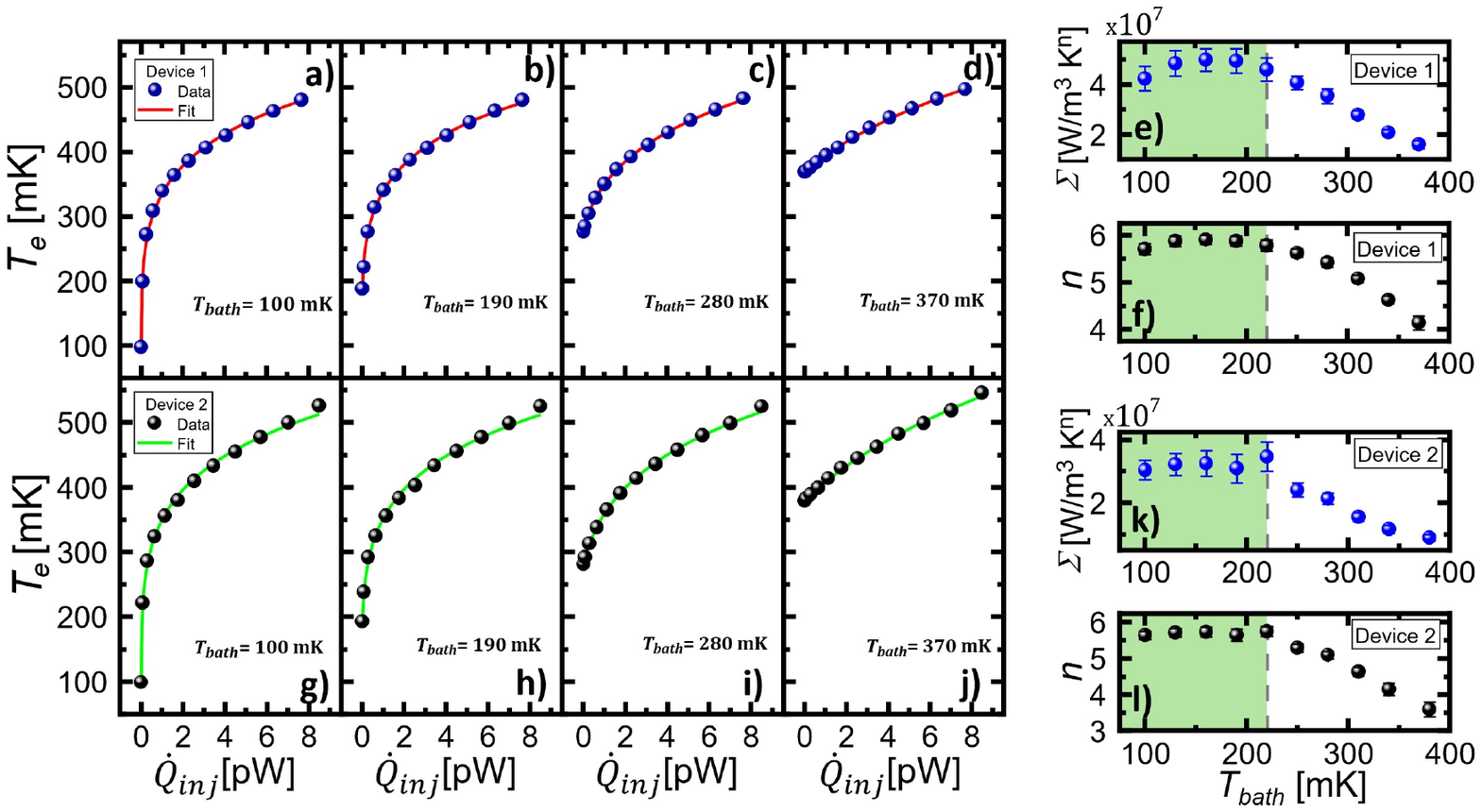}
    \caption{
    \textbf{(a)}-\textbf{(d)}: Experimental data and the fit using Eq.\ref{eq3}  are shown at selected bath temperatures (indicated in each panel) for the electronic temperature $T_e$ as a function of the injected power $\dot Q_{inj}$ for device $1$.
     \textbf{(e)},\textbf{(f)}: The value of $\Sigma$ and $n$ extracted from the fitting procedure as a function of $T_{\text{bath}}$ for device 1. \textbf{(g)}-\textbf{(j)}: Experimental data and the fit using Eq.\ref{eq3}  are shown at selected bath temperatures (indicated in each panel) for the electronic temperature $T_e$ as a function of the injected power $\dot Q_{inj}$ for device $2$. \textbf{(k)},\textbf{(l)}: The value of $\Sigma$ and $n$ extracted from the fitting procedure as a function of $T_{\text{bath}}$ for device 2.}
  \label{Fig3}
\end{figure*}
The inset displayed in Fig.\ref{Fig2}(a) shows the scheme of the predominant incoming and outgoing heat currents for the InAs mesa. The input power $\dot Q_{inj}$ is delivered by the Joule heaters, and the outgoing heat current $\dot Q_{e-ph}$ represents the power relaxing into the InAlAs insulating buffer layer due to electron-phonon interaction. 
We assume that the power leakage through the contacts (i.e., the Joule heaters, the JJ, and the other two unused contacts) on the InAs mesa is negligible because all those contacts are superconductors that behave, at low energy, as almost ideal Andreev mirrors. This is valid since the electron temperatures explored throughout all the measurements are in the limit $k_B T_e\ll \Delta_{Al}$, where $k_B$ is the Boltzmann constant and $\Delta_{Al}\simeq 180$ $\mu$eV \cite{tinkham2004}. 
In our analysis, we also assume that there is negligible electronic heat loss through the electrically insulating InAlAs buffer (with substrate resistance $>10$ T$\Omega$) and that the InAs lattice phonons are thermalized at the cryostat bath temperature $T_{bath}$ as well as all the rest of the heterostructure due to vanishing Kapitza resistance. Under these assumptions, we can write down the power balance equation for the InAs mesa in the stationary regime:
\begin{equation}
    \dot Q_{inj}-\dot Q_{e-ph}=0,
    \label{eq1}
\end{equation}
where $Q_{inj}=I_H V_H$, and $I_H$ and $V_H$ are the current and voltage drop across the Joule heaters in the InAs mesa. 
The dissipated power due to the e-ph interaction can be written as \cite{Wellstood1994,subero2023,gunnarsson2015,muhonen2011}:
\begin{equation}
    \dot Q_{e-ph}=\Sigma V (T_e^n - T_{bath}^n),
    \label{eq2}
\end{equation}
where $V$ represents the geometric volume of the InAs mesa, assumed to be uniformly thermalized, $\Sigma$ is the electron-phonon coupling constant that we aim to derive, and $n$ is the exponent of this power law. Typically,  $n=5$ for thin metallic films \cite{Wellstood1994}. 
However, both experimental evidence \cite{subero2023} and theoretical predictions \cite{sergeev2000,giazotto2006} indicate that this value can vary depending on the specific material and different regimes, such as ballistic or diffusive.
The electron-phonon heat losses of highly doped InAs have been studied since the 1970s \cite{bauer1972}.
Initially, it was predicted that the couplings would be very low.
Early estimations predicted that at extremely low temperatures (sub-Kelvin), the temperature dependencies would be very low and quasi-linear, approximately $n \approx 1$. 
However, a later study  \cite{price1982} found different values for $n$, specifically $n=5$ and $n=7$ for highly doped GaAs, related to piezoelectric coupling and deformation potential, respectively. 
The disorder can lead to changes in power laws in metals,  with values of $n$ falling between $4$ and $6$ \cite{sergeev2000}.
Due to these reasons, we opted to keep $n$ as a free-fitting parameter for the variability. By using  Eqs.\ref{eq1} and \ref{eq2}, we can express:
\begin{equation}
    T_e=\sqrt[n]{\frac{\dot Q_{inj}}{\Sigma V}+T_{bath}^n},
    \label{eq3}
\end{equation}
where $\Sigma$ and $n$ are fitting parameters of the dependence over $\dot Q_{inj}$ and $T_{bath}$.\\
By keeping the bath temperature $T_{bath}$ constant and adjusting the injected heat flux $\dot Q_{inj}$, we can measure the electronic temperature $T_e$ in the InAs mesa to extract the energy relaxation curve inside the InAs. 
Figures \ref{Fig3}(a-d) and (g-j) display selected curves for different bath temperatures for the two samples studied. 
These curves were then fitted with Eq.\ref{eq3}, and the results for the fitting parameters $\Sigma$ and $n$ are resumed in Figs.\ref{Fig3}(e,f) and (k,l). 
The agreement between the theory and the experiment 
is evident through the reduced chi-square $\chi_{red}^2$, with the worst value being approximately $\chi_{red}^2 \simeq 1.1\cdot 10^{-4}$.\\
In Fig.\ref{Fig3}(e,k), the electron-phonon coupling constant $\Sigma$, similarly to the exponent $n$, is shown to be almost constant in the green shaded region, ranging from $100$mK to $220$mK. The values are $\Sigma=(4.2\pm0.4)\cdot 10^7$ W/(m$^3$K$^n$) and $\Sigma=(3.0\pm0.3)\cdot 10^7$ W/(m$^3$K$^n$) for device 1 and 2, respectively. 
These values are notably low compared to those of other common metals \cite{giazotto2006} such as copper  [Cu, $\Sigma_{Cu}\simeq 2\cdot10^9$ W/(m$^3$K$^5$)] \cite{meschke2004}, aluminum [Al, $\Sigma_{Al}\simeq 0.2\cdot10^9$ W/(m$^3$K$^5$)] \cite{meschke2004}, aluminum manganese [AlMn, $\Sigma_{AlMn}\simeq4.5\cdot10^9$ W/(m$^3$K$^6$)] \cite{martinez-perez2014,martinez-perez2015,taskinen2006}, doped silicon on insulator [SOI, $\Sigma_{SOI}\simeq0.1\cdot10^9$ W/(m$^3$K$^5$)] \cite{savin2001}, and InAs nano-wires [$\Sigma_{InAsNW}\simeq4.7\cdot10^9$ W/(m$^3$K$^5$)] \cite{roddaro2011}.\\
When looking at Figs.\ref{Fig3}(f,l), we can see that the value of the exponent $n$ in the green shaded region from $100$mK to $220$mK remains almost constant ($n=5.84\pm0.14$ for device 1, $n=5.72\pm0.12$ for device 2). 
Once again, having a value of $n$ greater than $5$ means that phonon losses are even smaller at low temperatures, indicating that the InAsOI platform is well-suited for coherent caloritronics.\\
These properties mentioned above of InAsOI make it an excellent choice for caloritronic applications due to the minimal interaction between electrons and lattice phonons. The low electron-phonon coupling ensures that electronic temperature can be significantly altered with minimal power input, resulting in a more stable and controllable caloritronic technology with reduced phonon losses. The low values for the parameter $\Sigma$ compared to metals may be due to the lower electron carrier density. However, it is important to note that the main physical mechanism responsible for these properties has not been definitively identified. This is largely due to limited knowledge of other physical parameters required to calculate the electron-phonon coupling for this specific layered material. \\
\begin{figure}[t!]
  \includegraphics[width=1\columnwidth, trim= 2cm 0cm 1cm 0cm]{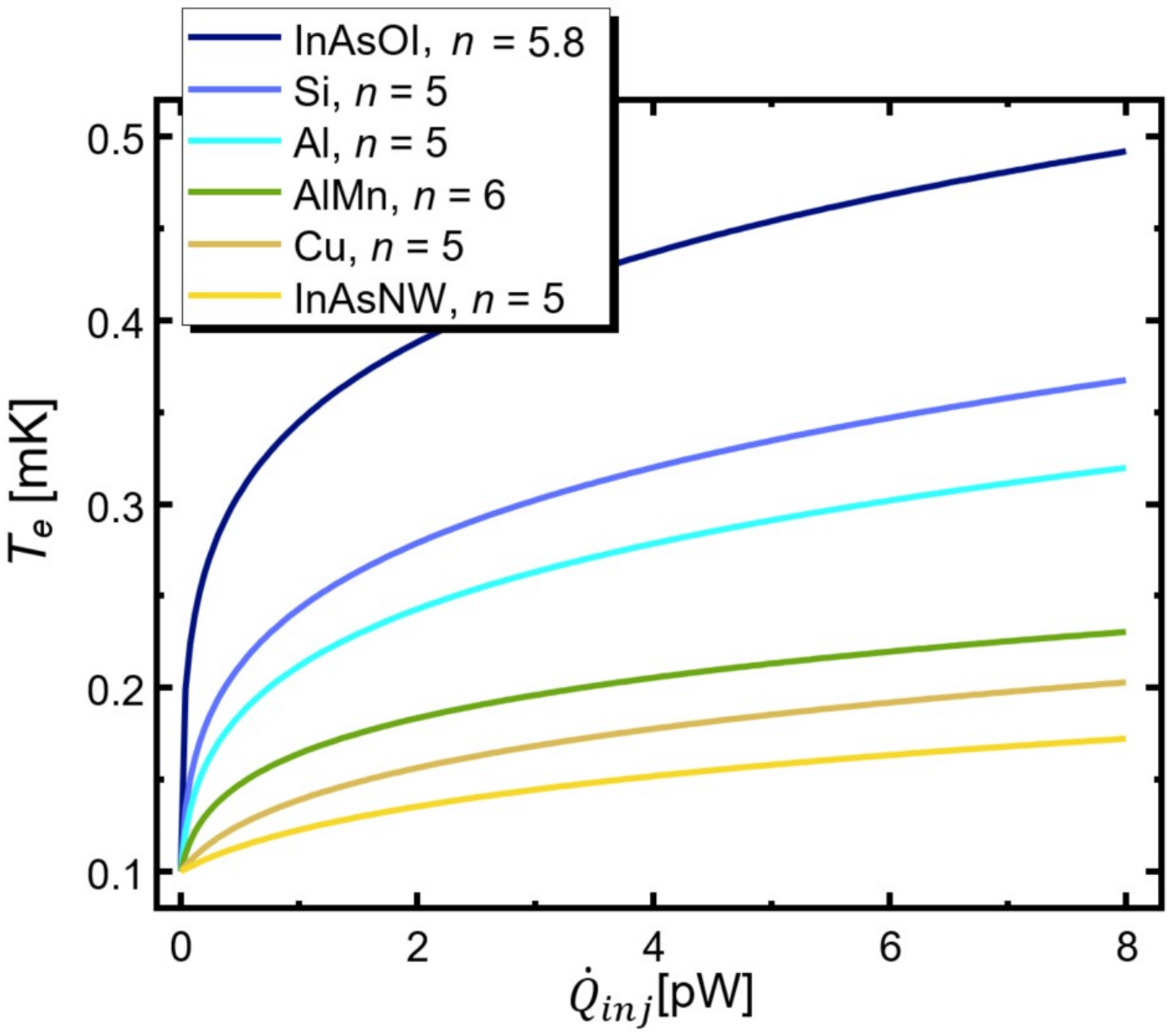}
    \caption{Comparison of the electronic temperature as a function of the injected power for different materials (as indicated in the legend) at a bath temperature of $T_{bath}=100$mK. The curve for InAsOI represents the fitted data obtained for device 1. In contrast, the curves for the other materials were calculated using literature values for $\Sigma$ and $n$ (SOI \cite{savin2001}, Al \cite{meschke2004}, AlMn \cite{martinez-perez2015}, Cu \cite{meschke2004}, and InAsNW \cite{roddaro2011}), applying  Eq.\ref{eq3} with the same volume as our InAs mesa.}
    \label{Fig4}
\end{figure}
 It is important to note that the electron-phonon coupling constants $\Sigma$ of different materials have different physical dimensions due to the different power law exponent $n$. 
 In Fig.\ref{Fig4}, we compare electronic temperature as a function of injected power for different materials.  The curve for our InAsOI device 1 at a bath temperature of $100$ mK is shown.  
For other materials, we calculated  theoretical curves using Eq.\ref{eq3}, with the same volume for the InAs mesa
 and $T_{bath}=100$ mK.
 This approach eliminates variations in physical dimensions for $\Sigma$ and $n$. It allows us to observe that the rise of $T_e$ in our InAsOI is significantly larger compared to other common materials when the same power is applied to the structure. 
 Although the change in electron temperature may seem small in absolute value, due to power-law scaling, it corresponds to differences in $\Sigma$ of orders of magnitude.  
 For example, comparing the lines for Al (cyan) with Cu (light brown), there is a difference of more than an order of magnitude in $\Sigma$. The expected electronic temperature variation for InAsOI is much higher than that of all other metallic materials. 
 This supports our conclusion that there is a very weak electron-phonon coupling, maintaining a good quality of the Josephson junction \cite{paghi2024}.\\
In the study of \text{III-V} materials at low temperatures (i.e., $T<1$ K), it was initially expected that the power exponent  $n$  would be 5 due to piezoelectric coupling \cite{price1982}. However, our research found that for $T_{bath}<220$ mK, the value of $n$ exceeded $5$, nearly reaching $6$.
This observation deviates from the predicted value due to certain conditions, specifically in the dirty limit, where the electron mean free path $l_e$ is much smaller than the phonon wavelength $\lambda_{ph}$, power dependencies like $T^6$ are predicted to occur \cite{sergeev2000}.
For our specific case, we calculated $l_e$ to be approximately $178$ nm using the equation $l_e=\mu v_F m^*/e$, where $v_F$ is the Fermi velocity, $m^*=0.023$ is the InAs effective mass, $\mu=4970$ cm$^2$/(V$\cdot$s) is the electron mobility, and $e$ is the electron charge. 
Additionally, we determined the phonon wavelength to be $\lambda_{ph}=h v_S/(k_B T_{bath})\simeq 1$ $\mu$m for $T_{bath}=200$ mK, where $h$ is the Planck constant, $k_B$ is the Boltzmann constant, and $v_S\simeq4.2\cdot 10^3$ m/s is the sound velocity in InAs \cite{degheidy2022}. 
These calculations indicate that our devices operate in the dirty limit where $l_e\ll \lambda_{ph}$, potentially contributing to the observed $n$ value exceeding  5. 
Nevertheless, the underlined reasons for the small $\Sigma$ and high $n$ values remain unclear.\\
The values of $\Sigma$ and $n$ show another unexpected feature: in Fig.\ref{Fig3}(e,f)-(k,l), it is evident that both parameters start to decay for $T_{bath}>220$ mK. This behavior can be explained by considering the electron-phonon scattering length, $L_{e-ph}$. This length can be estimated as $L_{e-ph}=\sqrt{D\tau_{e-ph}}$, where $\tau_{e-ph}$ is the electron-phonon scattering time calculated using the relation $\tau_{e-ph}=C/G_{th}$. Here,  $C=\gamma V T_e$ is the thermal capacitance of the electron gas, $\gamma$ is the Sommerfeld constant, where $\gamma=\pi^2 n_{3D} k_B/2T_F$,  $T_F=E_F/k_B$ is the Fermi temperature, $E_F=(\hbar^2/2m^*) (3\pi^2 n_{3D})^{2/3}$ is the Fermi energy, and $G_{th}=\partial \dot Q_{e-ph}/\partial T_e = n\Sigma V T_e^{n-1}$ is the thermal conductance of the electron gas. 
When we calculate $L_{e-ph}$ using these parameters we find:
\begin{equation}
    L_{e-ph}=\sqrt{\frac{\pi^2 k_B^2}{\hbar^2} \frac{m^*n_{3D}^{1/3}}{(3\pi^2)^{2/3}} \frac{v_F l_e}{3n \Sigma T_e^{n-2}}} .
\end{equation}
The $L_{e-ph}$ value measures the distance over which one can assume a constant electronic temperature. 
When using the values of $\Sigma$ and $n$ found in the $100$-$220$ mK region, this distance becomes shorter than $20\mu$m (which is the longitudinal dimension of the InAs mesa) when $T_e\simeq380$ mK. 
This could make our model unreliable for fitting the data at higher temperatures, and the decay of $\Sigma$ and $n$ could be ascribed to this.\\

In summary, we have conducted the fabrication and experimental analysis of a hybrid superconductor-semiconductor structure using InAs on insulator (InAsOI) and aluminum as a platform for caloritronic systems.  The performance of the electron-phonon coupling, compared to metals such as chromium (Cr) \cite{subero2023}, copper (Cu) \cite{giazotto2012} or gold-palladium (AuPd) \cite{meschke2006}, is remarkably good with values of $\Sigma\simeq3\cdot10^7$ W/(m$^3$K$^n$) and $n\simeq5.8$. 
Looking at other semiconductors \cite{price1982,roddaro2011} the comparison is encouraging due to the platform's low electron-phonon interaction and the potential to implement effects like superconducting proximitization and Josephson coupling, which are hardly feasible in materials such as Si \cite{buonomoE2003,gunnarsson2015}. 
This makes the platform an ideal candidate for all coherent caloritronics experiments \cite{fornieri2017,fornieri2015,hwang2020,subero2023}, as well as single-photon detection and bolometry \cite{giazotto2008,virtanen2018,kokkoniemi2020,kokkoniemi2019,solinas2018}. Furthermore, the semiconductive nature of InAsOI allows for modifying its properties, such as charge density, through external electrostatic gating. This ability, not possible with conventional superconducting metals, presents further opportunities for coherent caloritronics and gateable superconductivity.

\section{Aknowledgements}
We thank Marco Bernardi, Omer Arif, and Giacomo Trupiano for fruitful discussions.\\
SB, GdS, AB, AP, and FG acknowledge the EU’s Horizon 2020 Research
and Innovation Framework Programme under Grant
No. 964398 (SUPERGATE), No. 101057977 (SPECTRUM), and the PNRR MUR project PE0000023-NQSTI for partial financial support. AB acknowledges MUR-PRIN 2022 - Grant No. 2022B9P8LN-(PE3)-Project NEThEQS “Non-equilibrium coherent thermal effects in quantum systems” in PNRR Mission 4-Component 2-Investment 1.1 “Fondo per il Programma Nazionale di Ricerca e Progetti di Rilevante Interesse Nazionale (PRIN)” funded by the European Union-Next Generation EU, the Royal Society through the International Exchanges between the UK and Italy (Grants No. IEC R2 192166) and CNR project QTHERMONANO. LS acknowledges the National Infrastructure of the National Research Council (CNR) for quantum simulations and calculations “PASQUA".

\bibliography{Th_eval_InAs}

\begin{thebibliography}{51}%
\makeatletter
\providecommand \@ifxundefined [1]{%
 \@ifx{#1\undefined}
}%
\providecommand \@ifnum [1]{%
 \ifnum #1\expandafter \@firstoftwo
 \else \expandafter \@secondoftwo
 \fi
}%
\providecommand \@ifx [1]{%
 \ifx #1\expandafter \@firstoftwo
 \else \expandafter \@secondoftwo
 \fi
}%
\providecommand \natexlab [1]{#1}%
\providecommand \enquote  [1]{``#1''}%
\providecommand \bibnamefont  [1]{#1}%
\providecommand \bibfnamefont [1]{#1}%
\providecommand \citenamefont [1]{#1}%
\providecommand \href@noop [0]{\@secondoftwo}%
\providecommand \href [0]{\begingroup \@sanitize@url \@href}%
\providecommand \@href[1]{\@@startlink{#1}\@@href}%
\providecommand \@@href[1]{\endgroup#1\@@endlink}%
\providecommand \@sanitize@url [0]{\catcode `\\12\catcode `\$12\catcode `\&12\catcode `\#12\catcode `\^12\catcode `\_12\catcode `\%12\relax}%
\providecommand \@@startlink[1]{}%
\providecommand \@@endlink[0]{}%
\providecommand \url  [0]{\begingroup\@sanitize@url \@url }%
\providecommand \@url [1]{\endgroup\@href {#1}{\urlprefix }}%
\providecommand \urlprefix  [0]{URL }%
\providecommand \Eprint [0]{\href }%
\providecommand \doibase [0]{http://dx.doi.org/}%
\providecommand \selectlanguage [0]{\@gobble}%
\providecommand \bibinfo  [0]{\@secondoftwo}%
\providecommand \bibfield  [0]{\@secondoftwo}%
\providecommand \translation [1]{[#1]}%
\providecommand \BibitemOpen [0]{}%
\providecommand \bibitemStop [0]{}%
\providecommand \bibitemNoStop [0]{.\EOS\space}%
\providecommand \EOS [0]{\spacefactor3000\relax}%
\providecommand \BibitemShut  [1]{\csname bibitem#1\endcsname}%
\let\auto@bib@innerbib\@empty
%</preamble>
\bibitem [{\citenamefont {Fornieri}\ and\ \citenamefont {Giazotto}(2017)}]{fornieri2017}%
  \BibitemOpen
  \bibfield  {author} {\bibinfo {author} {\bibfnamefont {A.}~\bibnamefont {Fornieri}}\ and\ \bibinfo {author} {\bibfnamefont {F.}~\bibnamefont {Giazotto}},\ }\href {\doibase 10.1038/nnano.2017.204} {\bibfield  {journal} {\bibinfo  {journal} {Nat. Nanotechnol.}\ }\textbf {\bibinfo {volume} {12}},\ \bibinfo {pages} {944} (\bibinfo {year} {2017})}\BibitemShut {NoStop}%
\bibitem [{\citenamefont {Giazotto}\ \emph {et~al.}(2006)\citenamefont {Giazotto}, \citenamefont {Heikkil{\"a}}, \citenamefont {Luukanen}, \citenamefont {Savin},\ and\ \citenamefont {Pekola}}]{giazotto2006}%
  \BibitemOpen
  \bibfield  {author} {\bibinfo {author} {\bibfnamefont {F.}~\bibnamefont {Giazotto}}, \bibinfo {author} {\bibfnamefont {T.~T.}\ \bibnamefont {Heikkil{\"a}}}, \bibinfo {author} {\bibfnamefont {A.}~\bibnamefont {Luukanen}}, \bibinfo {author} {\bibfnamefont {A.~M.}\ \bibnamefont {Savin}}, \ and\ \bibinfo {author} {\bibfnamefont {J.~P.}\ \bibnamefont {Pekola}},\ }\href {\doibase 10.1103/RevModPhys.78.217} {\bibfield  {journal} {\bibinfo  {journal} {Rev. Mod. Phys.}\ }\textbf {\bibinfo {volume} {78}},\ \bibinfo {pages} {217} (\bibinfo {year} {2006})}\BibitemShut {NoStop}%
\bibitem [{\citenamefont {Hwang}\ and\ \citenamefont {Sothmann}(2020)}]{hwang2020}%
  \BibitemOpen
  \bibfield  {author} {\bibinfo {author} {\bibfnamefont {S.-Y.}\ \bibnamefont {Hwang}}\ and\ \bibinfo {author} {\bibfnamefont {B.}~\bibnamefont {Sothmann}},\ }\href {\doibase 10.1140/epjst/e2019-900094-y} {\bibfield  {journal} {\bibinfo  {journal} {Eur. Phys. J. Spec. Top.}\ }\textbf {\bibinfo {volume} {229}},\ \bibinfo {pages} {683} (\bibinfo {year} {2020})}\BibitemShut {NoStop}%
\bibitem [{\citenamefont {Linder}\ and\ \citenamefont {Bathen}(2016)}]{linder2016}%
  \BibitemOpen
  \bibfield  {author} {\bibinfo {author} {\bibfnamefont {J.}~\bibnamefont {Linder}}\ and\ \bibinfo {author} {\bibfnamefont {M.~E.}\ \bibnamefont {Bathen}},\ }\href {\doibase 10.1103/PhysRevB.93.224509} {\bibfield  {journal} {\bibinfo  {journal} {Phys. Rev. B}\ }\textbf {\bibinfo {volume} {93}},\ \bibinfo {pages} {224509} (\bibinfo {year} {2016})}\BibitemShut {NoStop}%
\bibitem [{\citenamefont {Wang}\ and\ \citenamefont {Li}(2008)}]{wang2008}%
  \BibitemOpen
  \bibfield  {author} {\bibinfo {author} {\bibfnamefont {L.}~\bibnamefont {Wang}}\ and\ \bibinfo {author} {\bibfnamefont {B.}~\bibnamefont {Li}},\ }\href {\doibase 10.1103/PhysRevLett.101.267203} {\bibfield  {journal} {\bibinfo  {journal} {Phys. Rev. Lett.}\ }\textbf {\bibinfo {volume} {101}},\ \bibinfo {pages} {267203} (\bibinfo {year} {2008})}\BibitemShut {NoStop}%
\bibitem [{\citenamefont {{Ordonez-Miranda}}\ \emph {et~al.}(2019)\citenamefont {{Ordonez-Miranda}}, \citenamefont {Ezzahri}, \citenamefont {{Tiburcio-Moreno}}, \citenamefont {Joulain},\ and\ \citenamefont {Drevillon}}]{ordonez-miranda2019}%
  \BibitemOpen
  \bibfield  {author} {\bibinfo {author} {\bibfnamefont {J.}~\bibnamefont {{Ordonez-Miranda}}}, \bibinfo {author} {\bibfnamefont {Y.}~\bibnamefont {Ezzahri}}, \bibinfo {author} {\bibfnamefont {J.~A.}\ \bibnamefont {{Tiburcio-Moreno}}}, \bibinfo {author} {\bibfnamefont {K.}~\bibnamefont {Joulain}}, \ and\ \bibinfo {author} {\bibfnamefont {J.}~\bibnamefont {Drevillon}},\ }\href {\doibase 10.1103/PhysRevLett.123.025901} {\bibfield  {journal} {\bibinfo  {journal} {Phys. Rev. Lett.}\ }\textbf {\bibinfo {volume} {123}},\ \bibinfo {pages} {025901} (\bibinfo {year} {2019})}\BibitemShut {NoStop}%
\bibitem [{\citenamefont {Kubytskyi}\ \emph {et~al.}(2014)\citenamefont {Kubytskyi}, \citenamefont {Biehs},\ and\ \citenamefont {{Ben-Abdallah}}}]{kubytskyi2014}%
  \BibitemOpen
  \bibfield  {author} {\bibinfo {author} {\bibfnamefont {V.}~\bibnamefont {Kubytskyi}}, \bibinfo {author} {\bibfnamefont {S.-A.}\ \bibnamefont {Biehs}}, \ and\ \bibinfo {author} {\bibfnamefont {P.}~\bibnamefont {{Ben-Abdallah}}},\ }\href {\doibase 10.1103/PhysRevLett.113.074301} {\bibfield  {journal} {\bibinfo  {journal} {Phys. Rev. Lett.}\ }\textbf {\bibinfo {volume} {113}},\ \bibinfo {pages} {074301} (\bibinfo {year} {2014})}\BibitemShut {NoStop}%
\bibitem [{\citenamefont {Germanese}\ \emph {et~al.}(2022)\citenamefont {Germanese}, \citenamefont {Paolucci}, \citenamefont {Marchegiani}, \citenamefont {Braggio},\ and\ \citenamefont {Giazotto}}]{germanese2022}%
  \BibitemOpen
  \bibfield  {author} {\bibinfo {author} {\bibfnamefont {G.}~\bibnamefont {Germanese}}, \bibinfo {author} {\bibfnamefont {F.}~\bibnamefont {Paolucci}}, \bibinfo {author} {\bibfnamefont {G.}~\bibnamefont {Marchegiani}}, \bibinfo {author} {\bibfnamefont {A.}~\bibnamefont {Braggio}}, \ and\ \bibinfo {author} {\bibfnamefont {F.}~\bibnamefont {Giazotto}},\ }\href@noop {} {\bibfield  {journal} {\bibinfo  {journal} {Nature Nanotech.}\ }\textbf {\bibinfo {volume} {17}},\ \bibinfo {pages} {1084} (\bibinfo {year} {2022})}\BibitemShut {NoStop}%
\bibitem [{\citenamefont {Fornieri}\ \emph {et~al.}(2015)\citenamefont {Fornieri}, \citenamefont {{Mart{\'i}nez-P{\'e}rez}},\ and\ \citenamefont {Giazotto}}]{fornieri2015}%
  \BibitemOpen
  \bibfield  {author} {\bibinfo {author} {\bibfnamefont {A.}~\bibnamefont {Fornieri}}, \bibinfo {author} {\bibfnamefont {M.~J.}\ \bibnamefont {{Mart{\'i}nez-P{\'e}rez}}}, \ and\ \bibinfo {author} {\bibfnamefont {F.}~\bibnamefont {Giazotto}},\ }\href {\doibase 10.1063/1.4915899} {\bibfield  {journal} {\bibinfo  {journal} {AIP Adv.}\ }\textbf {\bibinfo {volume} {5}},\ \bibinfo {pages} {053301} (\bibinfo {year} {2015})}\BibitemShut {NoStop}%
\bibitem [{\citenamefont {Li}\ \emph {et~al.}(2004)\citenamefont {Li}, \citenamefont {Wang},\ and\ \citenamefont {Casati}}]{li2004}%
  \BibitemOpen
  \bibfield  {author} {\bibinfo {author} {\bibfnamefont {B.}~\bibnamefont {Li}}, \bibinfo {author} {\bibfnamefont {L.}~\bibnamefont {Wang}}, \ and\ \bibinfo {author} {\bibfnamefont {G.}~\bibnamefont {Casati}},\ }\href {\doibase 10.1103/PhysRevLett.93.184301} {\bibfield  {journal} {\bibinfo  {journal} {Phys. Rev. Lett.}\ }\textbf {\bibinfo {volume} {93}},\ \bibinfo {pages} {184301} (\bibinfo {year} {2004})}\BibitemShut {NoStop}%
\bibitem [{\citenamefont {Paolucci}\ \emph {et~al.}(2017)\citenamefont {Paolucci}, \citenamefont {Marchegiani}, \citenamefont {Strambini},\ and\ \citenamefont {Giazotto}}]{paolucci2017}%
  \BibitemOpen
  \bibfield  {author} {\bibinfo {author} {\bibfnamefont {F.}~\bibnamefont {Paolucci}}, \bibinfo {author} {\bibfnamefont {G.}~\bibnamefont {Marchegiani}}, \bibinfo {author} {\bibfnamefont {E.}~\bibnamefont {Strambini}}, \ and\ \bibinfo {author} {\bibfnamefont {F.}~\bibnamefont {Giazotto}},\ }\href {\doibase 10.1209/0295-5075/118/68004} {\bibfield  {journal} {\bibinfo  {journal} {Europhys. Lett.}\ }\textbf {\bibinfo {volume} {118}},\ \bibinfo {pages} {68004} (\bibinfo {year} {2017})}\BibitemShut {NoStop}%
\bibitem [{\citenamefont {Wang}\ and\ \citenamefont {Li}(2007)}]{wang2007}%
  \BibitemOpen
  \bibfield  {author} {\bibinfo {author} {\bibfnamefont {L.}~\bibnamefont {Wang}}\ and\ \bibinfo {author} {\bibfnamefont {B.}~\bibnamefont {Li}},\ }\href {\doibase 10.1103/PhysRevLett.99.177208} {\bibfield  {journal} {\bibinfo  {journal} {Phys. Rev. Lett.}\ }\textbf {\bibinfo {volume} {99}},\ \bibinfo {pages} {177208} (\bibinfo {year} {2007})}\BibitemShut {NoStop}%
\bibitem [{\citenamefont {Paolucci}\ \emph {et~al.}(2018)\citenamefont {Paolucci}, \citenamefont {Marchegiani}, \citenamefont {Strambini},\ and\ \citenamefont {Giazotto}}]{paolucci2018}%
  \BibitemOpen
  \bibfield  {author} {\bibinfo {author} {\bibfnamefont {F.}~\bibnamefont {Paolucci}}, \bibinfo {author} {\bibfnamefont {G.}~\bibnamefont {Marchegiani}}, \bibinfo {author} {\bibfnamefont {E.}~\bibnamefont {Strambini}}, \ and\ \bibinfo {author} {\bibfnamefont {F.}~\bibnamefont {Giazotto}},\ }\href {\doibase 10.1103/PhysRevApplied.10.024003} {\bibfield  {journal} {\bibinfo  {journal} {Phys. Rev. Appl.}\ }\textbf {\bibinfo {volume} {10}},\ \bibinfo {pages} {024003} (\bibinfo {year} {2018})}\BibitemShut {NoStop}%
\bibitem [{\citenamefont {Liu}\ \emph {et~al.}(2024)\citenamefont {Liu}, \citenamefont {Mao},\ and\ \citenamefont {Sun}}]{liu2024}%
  \BibitemOpen
  \bibfield  {author} {\bibinfo {author} {\bibfnamefont {P.-Y.}\ \bibnamefont {Liu}}, \bibinfo {author} {\bibfnamefont {Y.}~\bibnamefont {Mao}}, \ and\ \bibinfo {author} {\bibfnamefont {Q.-F.}\ \bibnamefont {Sun}},\ }\href {\doibase 10.1103/PhysRevApplied.21.024001} {\bibfield  {journal} {\bibinfo  {journal} {Phys. Rev. Appl.}\ }\textbf {\bibinfo {volume} {21}},\ \bibinfo {pages} {024001} (\bibinfo {year} {2024})}\BibitemShut {NoStop}%
\bibitem [{\citenamefont {Bauer}\ and\ \citenamefont {Kahlert}(1972)}]{bauer1972}%
  \BibitemOpen
  \bibfield  {author} {\bibinfo {author} {\bibfnamefont {G.}~\bibnamefont {Bauer}}\ and\ \bibinfo {author} {\bibfnamefont {H.}~\bibnamefont {Kahlert}},\ }\href {\doibase 10.1103/PhysRevB.5.566} {\bibfield  {journal} {\bibinfo  {journal} {Phys. Rev. B}\ }\textbf {\bibinfo {volume} {5}},\ \bibinfo {pages} {566} (\bibinfo {year} {1972})}\BibitemShut {NoStop}%
\bibitem [{\citenamefont {Price}(1982)}]{price1982}%
  \BibitemOpen
  \bibfield  {author} {\bibinfo {author} {\bibfnamefont {P.~J.}\ \bibnamefont {Price}},\ }\href {\doibase 10.1063/1.330026} {\bibfield  {journal} {\bibinfo  {journal} {J. Appl. Phys.}\ }\textbf {\bibinfo {volume} {53}},\ \bibinfo {pages} {6863} (\bibinfo {year} {1982})}\BibitemShut {NoStop}%
\bibitem [{\citenamefont {Muhonen}\ \emph {et~al.}(2011)\citenamefont {Muhonen}, \citenamefont {Prest}, \citenamefont {Prunnila}, \citenamefont {Gunnarsson}, \citenamefont {Shah}, \citenamefont {Dobbie}, \citenamefont {Myronov}, \citenamefont {Morris}, \citenamefont {Whall}, \citenamefont {Parker},\ and\ \citenamefont {Leadley}}]{muhonen2011}%
  \BibitemOpen
  \bibfield  {author} {\bibinfo {author} {\bibfnamefont {J.~T.}\ \bibnamefont {Muhonen}}, \bibinfo {author} {\bibfnamefont {M.~J.}\ \bibnamefont {Prest}}, \bibinfo {author} {\bibfnamefont {M.}~\bibnamefont {Prunnila}}, \bibinfo {author} {\bibfnamefont {D.}~\bibnamefont {Gunnarsson}}, \bibinfo {author} {\bibfnamefont {V.~A.}\ \bibnamefont {Shah}}, \bibinfo {author} {\bibfnamefont {A.}~\bibnamefont {Dobbie}}, \bibinfo {author} {\bibfnamefont {M.}~\bibnamefont {Myronov}}, \bibinfo {author} {\bibfnamefont {R.~J.~H.}\ \bibnamefont {Morris}}, \bibinfo {author} {\bibfnamefont {T.~E.}\ \bibnamefont {Whall}}, \bibinfo {author} {\bibfnamefont {E.~H.~C.}\ \bibnamefont {Parker}}, \ and\ \bibinfo {author} {\bibfnamefont {D.~R.}\ \bibnamefont {Leadley}},\ }\href {\doibase 10.1063/1.3579524} {\bibfield  {journal} {\bibinfo  {journal} {Appl. Phys. Lett.}\ }\textbf {\bibinfo {volume} {98}},\ \bibinfo {pages} {182103} (\bibinfo {year} {2011})}\BibitemShut {NoStop}%
\bibitem [{Note1()}]{Note1}%
  \BibitemOpen
  \bibinfo {note} {F. Giazotto, G. De Simoni, L. Sorba, A. Paghi, O. Arif, and C. Puglia, ''Josephson Junction and superconductor field effect transistor'', Filling number: 102024000008983 (19/04/2024)}\BibitemShut {NoStop}%
\bibitem [{\citenamefont {Paghi}\ \emph {et~al.}(2024)\citenamefont {Paghi}, \citenamefont {Trupiano}, \citenamefont {De~Simoni}, \citenamefont {Arif}, \citenamefont {Sorba},\ and\ \citenamefont {Giazotto}}]{paghi2024}%
  \BibitemOpen
  \bibfield  {author} {\bibinfo {author} {\bibfnamefont {A.}~\bibnamefont {Paghi}}, \bibinfo {author} {\bibfnamefont {G.}~\bibnamefont {Trupiano}}, \bibinfo {author} {\bibfnamefont {G.}~\bibnamefont {De~Simoni}}, \bibinfo {author} {\bibfnamefont {O.}~\bibnamefont {Arif}}, \bibinfo {author} {\bibfnamefont {L.}~\bibnamefont {Sorba}}, \ and\ \bibinfo {author} {\bibfnamefont {F.}~\bibnamefont {Giazotto}},\ }\href@noop {} {\enquote {\bibinfo {title} {{{InAs}} on {{Insulator}}: {{A New Platform}} for {{Cryogenic Hybrid Superconducting Electronics}}},}\ } (\bibinfo {year} {2024}),\ \Eprint {http://arxiv.org/abs/2405.07630} {arxiv:2405.07630} \BibitemShut {NoStop}%
\bibitem [{\citenamefont {Wellstood}\ \emph {et~al.}(1994)\citenamefont {Wellstood}, \citenamefont {Urbina},\ and\ \citenamefont {Clarke}}]{Wellstood1994}%
  \BibitemOpen
  \bibfield  {author} {\bibinfo {author} {\bibfnamefont {F.~C.}\ \bibnamefont {Wellstood}}, \bibinfo {author} {\bibfnamefont {C.}~\bibnamefont {Urbina}}, \ and\ \bibinfo {author} {\bibfnamefont {J.}~\bibnamefont {Clarke}},\ }\href {\doibase 10.1103/PhysRevB.49.5942} {\bibfield  {journal} {\bibinfo  {journal} {Phys. Rev. B}\ }\textbf {\bibinfo {volume} {49}},\ \bibinfo {pages} {5942} (\bibinfo {year} {1994})}\BibitemShut {NoStop}%
\bibitem [{\citenamefont {Arif}\ \emph {et~al.}(2024)\citenamefont {Arif}, \citenamefont {Canal}, \citenamefont {Ferrari}, \citenamefont {Ferrari}, \citenamefont {Lazzarini}, \citenamefont {Nasi}, \citenamefont {Paghi}, \citenamefont {Heun},\ and\ \citenamefont {Sorba}}]{arif2024}%
  \BibitemOpen
  \bibfield  {author} {\bibinfo {author} {\bibfnamefont {O.}~\bibnamefont {Arif}}, \bibinfo {author} {\bibfnamefont {L.}~\bibnamefont {Canal}}, \bibinfo {author} {\bibfnamefont {E.}~\bibnamefont {Ferrari}}, \bibinfo {author} {\bibfnamefont {C.}~\bibnamefont {Ferrari}}, \bibinfo {author} {\bibfnamefont {L.}~\bibnamefont {Lazzarini}}, \bibinfo {author} {\bibfnamefont {L.}~\bibnamefont {Nasi}}, \bibinfo {author} {\bibfnamefont {A.}~\bibnamefont {Paghi}}, \bibinfo {author} {\bibfnamefont {S.}~\bibnamefont {Heun}}, \ and\ \bibinfo {author} {\bibfnamefont {L.}~\bibnamefont {Sorba}},\ }\href {\doibase 10.3390/nano14070592} {\bibfield  {journal} {\bibinfo  {journal} {Nanomat.}\ }\textbf {\bibinfo {volume} {14}},\ \bibinfo {pages} {592} (\bibinfo {year} {2024})}\BibitemShut {NoStop}%
\bibitem [{\citenamefont {Benali}\ \emph {et~al.}(2022)\citenamefont {Benali}, \citenamefont {Rajak}, \citenamefont {Ciancio}, \citenamefont {Plaisier}, \citenamefont {Heun},\ and\ \citenamefont {Biasiol}}]{benali2022}%
  \BibitemOpen
  \bibfield  {author} {\bibinfo {author} {\bibfnamefont {A.}~\bibnamefont {Benali}}, \bibinfo {author} {\bibfnamefont {P.}~\bibnamefont {Rajak}}, \bibinfo {author} {\bibfnamefont {R.}~\bibnamefont {Ciancio}}, \bibinfo {author} {\bibfnamefont {J.~R.}\ \bibnamefont {Plaisier}}, \bibinfo {author} {\bibfnamefont {S.}~\bibnamefont {Heun}}, \ and\ \bibinfo {author} {\bibfnamefont {G.}~\bibnamefont {Biasiol}},\ }\href {\doibase 10.1016/j.jcrysgro.2022.126768} {\bibfield  {journal} {\bibinfo  {journal} {J. Cryst. Growth}\ }\textbf {\bibinfo {volume} {593}},\ \bibinfo {pages} {126768} (\bibinfo {year} {2022})}\BibitemShut {NoStop}%
\bibitem [{\citenamefont {Giazotto}\ \emph {et~al.}(2001{\natexlab{a}})\citenamefont {Giazotto}, \citenamefont {Cecchini}, \citenamefont {Pingue}, \citenamefont {Beltram}, \citenamefont {Lazzarino}, \citenamefont {Orani}, \citenamefont {Rubini},\ and\ \citenamefont {Franciosi}}]{giazotto2001}%
  \BibitemOpen
  \bibfield  {author} {\bibinfo {author} {\bibfnamefont {F.}~\bibnamefont {Giazotto}}, \bibinfo {author} {\bibfnamefont {M.}~\bibnamefont {Cecchini}}, \bibinfo {author} {\bibfnamefont {P.}~\bibnamefont {Pingue}}, \bibinfo {author} {\bibfnamefont {F.}~\bibnamefont {Beltram}}, \bibinfo {author} {\bibfnamefont {M.}~\bibnamefont {Lazzarino}}, \bibinfo {author} {\bibfnamefont {D.}~\bibnamefont {Orani}}, \bibinfo {author} {\bibfnamefont {S.}~\bibnamefont {Rubini}}, \ and\ \bibinfo {author} {\bibfnamefont {A.}~\bibnamefont {Franciosi}},\ }\href {\doibase 10.1063/1.1357211} {\bibfield  {journal} {\bibinfo  {journal} {Appl. Phys. Lett.}\ }\textbf {\bibinfo {volume} {78}},\ \bibinfo {pages} {1772} (\bibinfo {year} {2001}{\natexlab{a}})}\BibitemShut {NoStop}%
\bibitem [{\citenamefont {Giazotto}\ \emph {et~al.}(2001{\natexlab{b}})\citenamefont {Giazotto}, \citenamefont {Pingue}, \citenamefont {Beltram}, \citenamefont {Lazzarino}, \citenamefont {Orani}, \citenamefont {Rubini},\ and\ \citenamefont {Franciosi}}]{giazotto2001b}%
  \BibitemOpen
  \bibfield  {author} {\bibinfo {author} {\bibfnamefont {F.}~\bibnamefont {Giazotto}}, \bibinfo {author} {\bibfnamefont {P.}~\bibnamefont {Pingue}}, \bibinfo {author} {\bibfnamefont {F.}~\bibnamefont {Beltram}}, \bibinfo {author} {\bibfnamefont {M.}~\bibnamefont {Lazzarino}}, \bibinfo {author} {\bibfnamefont {D.}~\bibnamefont {Orani}}, \bibinfo {author} {\bibfnamefont {S.}~\bibnamefont {Rubini}}, \ and\ \bibinfo {author} {\bibfnamefont {A.}~\bibnamefont {Franciosi}},\ }\href {\doibase 10.1103/PhysRevLett.87.216808} {\bibfield  {journal} {\bibinfo  {journal} {Appl. Phys. Lett.}\ }\textbf {\bibinfo {volume} {87}},\ \bibinfo {pages} {216808} (\bibinfo {year} {2001}{\natexlab{b}})}\BibitemShut {NoStop}%
\bibitem [{\citenamefont {Cochran}\ and\ \citenamefont {Mapother}(1958)}]{cochran1958}%
  \BibitemOpen
  \bibfield  {author} {\bibinfo {author} {\bibfnamefont {J.~F.}\ \bibnamefont {Cochran}}\ and\ \bibinfo {author} {\bibfnamefont {D.~E.}\ \bibnamefont {Mapother}},\ }\href {\doibase 10.1103/PhysRev.111.132} {\bibfield  {journal} {\bibinfo  {journal} {Phys. Rev.}\ }\textbf {\bibinfo {volume} {111}},\ \bibinfo {pages} {132} (\bibinfo {year} {1958})}\BibitemShut {NoStop}%
\bibitem [{\citenamefont {Bhargava}\ \emph {et~al.}(1997)\citenamefont {Bhargava}, \citenamefont {Blank}, \citenamefont {Narayanamurti},\ and\ \citenamefont {Kroemer}}]{bhargava1997}%
  \BibitemOpen
  \bibfield  {author} {\bibinfo {author} {\bibfnamefont {S.}~\bibnamefont {Bhargava}}, \bibinfo {author} {\bibfnamefont {H.-R.}\ \bibnamefont {Blank}}, \bibinfo {author} {\bibfnamefont {V.}~\bibnamefont {Narayanamurti}}, \ and\ \bibinfo {author} {\bibfnamefont {H.}~\bibnamefont {Kroemer}},\ }\href {\doibase 10.1063/1.118271} {\bibfield  {journal} {\bibinfo  {journal} {Appl. Phys. Lett.}\ }\textbf {\bibinfo {volume} {70}},\ \bibinfo {pages} {759} (\bibinfo {year} {1997})}\BibitemShut {NoStop}%
\bibitem [{\citenamefont {Melitz}\ \emph {et~al.}(2010)\citenamefont {Melitz}, \citenamefont {Shen}, \citenamefont {Lee}, \citenamefont {Lee}, \citenamefont {Kummel}, \citenamefont {Droopad},\ and\ \citenamefont {Yu}}]{melitz2010}%
  \BibitemOpen
  \bibfield  {author} {\bibinfo {author} {\bibfnamefont {W.}~\bibnamefont {Melitz}}, \bibinfo {author} {\bibfnamefont {J.}~\bibnamefont {Shen}}, \bibinfo {author} {\bibfnamefont {S.}~\bibnamefont {Lee}}, \bibinfo {author} {\bibfnamefont {J.~S.}\ \bibnamefont {Lee}}, \bibinfo {author} {\bibfnamefont {A.~C.}\ \bibnamefont {Kummel}}, \bibinfo {author} {\bibfnamefont {R.}~\bibnamefont {Droopad}}, \ and\ \bibinfo {author} {\bibfnamefont {E.~T.}\ \bibnamefont {Yu}},\ }\href {\doibase 10.1063/1.3462440} {\bibfield  {journal} {\bibinfo  {journal} {J. Appl. Phys.}\ }\textbf {\bibinfo {volume} {108}},\ \bibinfo {pages} {023711} (\bibinfo {year} {2010})}\BibitemShut {NoStop}%
\bibitem [{\citenamefont {Dubos}\ \emph {et~al.}(2001)\citenamefont {Dubos}, \citenamefont {Courtois}, \citenamefont {Pannetier}, \citenamefont {Wilhelm}, \citenamefont {Zaikin},\ and\ \citenamefont {Sch{\"o}n}}]{dubos2001}%
  \BibitemOpen
  \bibfield  {author} {\bibinfo {author} {\bibfnamefont {P.}~\bibnamefont {Dubos}}, \bibinfo {author} {\bibfnamefont {H.}~\bibnamefont {Courtois}}, \bibinfo {author} {\bibfnamefont {B.}~\bibnamefont {Pannetier}}, \bibinfo {author} {\bibfnamefont {F.~K.}\ \bibnamefont {Wilhelm}}, \bibinfo {author} {\bibfnamefont {A.~D.}\ \bibnamefont {Zaikin}}, \ and\ \bibinfo {author} {\bibfnamefont {G.}~\bibnamefont {Sch{\"o}n}},\ }\href {\doibase 10.1103/PhysRevB.63.064502} {\bibfield  {journal} {\bibinfo  {journal} {Phys. Rev. B}\ }\textbf {\bibinfo {volume} {63}},\ \bibinfo {pages} {064502} (\bibinfo {year} {2001})}\BibitemShut {NoStop}%
\bibitem [{\citenamefont {Edwards}\ and\ \citenamefont {Thouless}(1972)}]{edwards1972}%
  \BibitemOpen
  \bibfield  {author} {\bibinfo {author} {\bibfnamefont {J.~T.}\ \bibnamefont {Edwards}}\ and\ \bibinfo {author} {\bibfnamefont {D.~J.}\ \bibnamefont {Thouless}},\ }\href {\doibase 10.1088/0022-3719/5/8/007} {\bibfield  {journal} {\bibinfo  {journal} {J. Phys. C: Solid State Phys.}\ }\textbf {\bibinfo {volume} {5}},\ \bibinfo {pages} {807} (\bibinfo {year} {1972})}\BibitemShut {NoStop}%
\bibitem [{\citenamefont {Barone}\ and\ \citenamefont {Paternò}(1982)}]{baronepaternò}%
  \BibitemOpen
  \bibfield  {author} {\bibinfo {author} {\bibfnamefont {A.}~\bibnamefont {Barone}}\ and\ \bibinfo {author} {\bibfnamefont {G.}~\bibnamefont {Paternò}},\ }\href@noop {} {\emph {\bibinfo {title} {Physics and Applications of the Josephson Effect}}}\ (\bibinfo  {publisher} {John Wiley \& Sons Inc},\ \bibinfo {year} {1982})\BibitemShut {NoStop}%
\bibitem [{\citenamefont {Suominen}\ \emph {et~al.}(2017)\citenamefont {Suominen}, \citenamefont {Danon}, \citenamefont {Kjaergaard}, \citenamefont {Flensberg}, \citenamefont {Shabani}, \citenamefont {Palmstr{\o}m}, \citenamefont {Nichele},\ and\ \citenamefont {Marcus}}]{suominen2017}%
  \BibitemOpen
  \bibfield  {author} {\bibinfo {author} {\bibfnamefont {H.~J.}\ \bibnamefont {Suominen}}, \bibinfo {author} {\bibfnamefont {J.}~\bibnamefont {Danon}}, \bibinfo {author} {\bibfnamefont {M.}~\bibnamefont {Kjaergaard}}, \bibinfo {author} {\bibfnamefont {K.}~\bibnamefont {Flensberg}}, \bibinfo {author} {\bibfnamefont {J.}~\bibnamefont {Shabani}}, \bibinfo {author} {\bibfnamefont {C.~J.}\ \bibnamefont {Palmstr{\o}m}}, \bibinfo {author} {\bibfnamefont {F.}~\bibnamefont {Nichele}}, \ and\ \bibinfo {author} {\bibfnamefont {C.~M.}\ \bibnamefont {Marcus}},\ }\href {\doibase 10.1103/PhysRevB.95.035307} {\bibfield  {journal} {\bibinfo  {journal} {Phys. Rev. B}\ }\textbf {\bibinfo {volume} {95}},\ \bibinfo {pages} {035307} (\bibinfo {year} {2017})}\BibitemShut {NoStop}%
\bibitem [{\citenamefont {Gasparinetti}\ \emph {et~al.}(2015)\citenamefont {Gasparinetti}, \citenamefont {Viisanen}, \citenamefont {Saira}, \citenamefont {Faivre}, \citenamefont {Arzeo}, \citenamefont {Meschke},\ and\ \citenamefont {Pekola}}]{gasparinetti2015}%
  \BibitemOpen
  \bibfield  {author} {\bibinfo {author} {\bibfnamefont {S.}~\bibnamefont {Gasparinetti}}, \bibinfo {author} {\bibfnamefont {K.~L.}\ \bibnamefont {Viisanen}}, \bibinfo {author} {\bibfnamefont {O.~P.}\ \bibnamefont {Saira}}, \bibinfo {author} {\bibfnamefont {T.}~\bibnamefont {Faivre}}, \bibinfo {author} {\bibfnamefont {M.}~\bibnamefont {Arzeo}}, \bibinfo {author} {\bibfnamefont {M.}~\bibnamefont {Meschke}}, \ and\ \bibinfo {author} {\bibfnamefont {J.}~\bibnamefont {Pekola}},\ }\href {\doibase https://doi.org/10.1103/PhysRevApplied.3.014007} {\bibfield  {journal} {\bibinfo  {journal} {Phys. Rev. Appl.}\ }\textbf {\bibinfo {volume} {3}},\ \bibinfo {pages} {014007} (\bibinfo {year} {2015})}\BibitemShut {NoStop}%
\bibitem [{\citenamefont {Virtanen}\ \emph {et~al.}(2018)\citenamefont {Virtanen}, \citenamefont {Ronzani},\ and\ \citenamefont {Giazotto}}]{virtanen2018}%
  \BibitemOpen
  \bibfield  {author} {\bibinfo {author} {\bibfnamefont {P.}~\bibnamefont {Virtanen}}, \bibinfo {author} {\bibfnamefont {A.}~\bibnamefont {Ronzani}}, \ and\ \bibinfo {author} {\bibfnamefont {F.}~\bibnamefont {Giazotto}},\ }\href {\doibase 10.1103/PhysRevApplied.9.054027} {\bibfield  {journal} {\bibinfo  {journal} {Phys. Rev. Appl.}\ }\textbf {\bibinfo {volume} {9}},\ \bibinfo {pages} {054027} (\bibinfo {year} {2018})}\BibitemShut {NoStop}%
\bibitem [{\citenamefont {Kokkoniemi}\ \emph {et~al.}(2020)\citenamefont {Kokkoniemi}, \citenamefont {Girard}, \citenamefont {Hazra}, \citenamefont {Laitinen}, \citenamefont {Govenius}, \citenamefont {Lake}, \citenamefont {Sallinen}, \citenamefont {Vesterinen}, \citenamefont {Partanen}, \citenamefont {Tan}, \citenamefont {Chan}, \citenamefont {Tan}, \citenamefont {Hakonen},\ and\ \citenamefont {M{\"o}tt{\"o}nen}}]{kokkoniemi2020}%
  \BibitemOpen
  \bibfield  {author} {\bibinfo {author} {\bibfnamefont {R.}~\bibnamefont {Kokkoniemi}}, \bibinfo {author} {\bibfnamefont {J.-P.}\ \bibnamefont {Girard}}, \bibinfo {author} {\bibfnamefont {D.}~\bibnamefont {Hazra}}, \bibinfo {author} {\bibfnamefont {A.}~\bibnamefont {Laitinen}}, \bibinfo {author} {\bibfnamefont {J.}~\bibnamefont {Govenius}}, \bibinfo {author} {\bibfnamefont {R.~E.}\ \bibnamefont {Lake}}, \bibinfo {author} {\bibfnamefont {I.}~\bibnamefont {Sallinen}}, \bibinfo {author} {\bibfnamefont {V.}~\bibnamefont {Vesterinen}}, \bibinfo {author} {\bibfnamefont {M.}~\bibnamefont {Partanen}}, \bibinfo {author} {\bibfnamefont {J.~Y.}\ \bibnamefont {Tan}}, \bibinfo {author} {\bibfnamefont {K.~W.}\ \bibnamefont {Chan}}, \bibinfo {author} {\bibfnamefont {K.~Y.}\ \bibnamefont {Tan}}, \bibinfo {author} {\bibfnamefont {P.}~\bibnamefont {Hakonen}}, \ and\ \bibinfo {author} {\bibfnamefont {M.}~\bibnamefont {M{\"o}tt{\"o}nen}},\ }\href {\doibase 10.1038/s41586-020-2753-3} {\bibfield  {journal} {\bibinfo  {journal}
  {Nature}\ }\textbf {\bibinfo {volume} {586}},\ \bibinfo {pages} {47} (\bibinfo {year} {2020})}\BibitemShut {NoStop}%
\bibitem [{\citenamefont {Kokkoniemi}\ \emph {et~al.}(2019)\citenamefont {Kokkoniemi}, \citenamefont {Govenius}, \citenamefont {Vesterinen}, \citenamefont {Lake}, \citenamefont {Gunyh{\'o}}, \citenamefont {Tan}, \citenamefont {Simbierowicz}, \citenamefont {Gr{\"o}nberg}, \citenamefont {Lehtinen}, \citenamefont {Prunnila}, \citenamefont {Hassel}, \citenamefont {Lamminen}, \citenamefont {Saira},\ and\ \citenamefont {M{\"o}tt{\"o}nen}}]{kokkoniemi2019}%
  \BibitemOpen
  \bibfield  {author} {\bibinfo {author} {\bibfnamefont {R.}~\bibnamefont {Kokkoniemi}}, \bibinfo {author} {\bibfnamefont {J.}~\bibnamefont {Govenius}}, \bibinfo {author} {\bibfnamefont {V.}~\bibnamefont {Vesterinen}}, \bibinfo {author} {\bibfnamefont {R.~E.}\ \bibnamefont {Lake}}, \bibinfo {author} {\bibfnamefont {A.~M.}\ \bibnamefont {Gunyh{\'o}}}, \bibinfo {author} {\bibfnamefont {K.~Y.}\ \bibnamefont {Tan}}, \bibinfo {author} {\bibfnamefont {S.}~\bibnamefont {Simbierowicz}}, \bibinfo {author} {\bibfnamefont {L.}~\bibnamefont {Gr{\"o}nberg}}, \bibinfo {author} {\bibfnamefont {J.}~\bibnamefont {Lehtinen}}, \bibinfo {author} {\bibfnamefont {M.}~\bibnamefont {Prunnila}}, \bibinfo {author} {\bibfnamefont {J.}~\bibnamefont {Hassel}}, \bibinfo {author} {\bibfnamefont {A.}~\bibnamefont {Lamminen}}, \bibinfo {author} {\bibfnamefont {O.-P.}\ \bibnamefont {Saira}}, \ and\ \bibinfo {author} {\bibfnamefont {M.}~\bibnamefont {M{\"o}tt{\"o}nen}},\ }\href {\doibase 10.1038/s42005-019-0225-6} {\bibfield  {journal}
  {\bibinfo  {journal} {Commun. Phys.}\ }\textbf {\bibinfo {volume} {2}},\ \bibinfo {pages} {1} (\bibinfo {year} {2019})}\BibitemShut {NoStop}%
\bibitem [{\citenamefont {Solinas}\ \emph {et~al.}(2018)\citenamefont {Solinas}, \citenamefont {Giazotto},\ and\ \citenamefont {Pepe}}]{solinas2018}%
  \BibitemOpen
  \bibfield  {author} {\bibinfo {author} {\bibfnamefont {P.}~\bibnamefont {Solinas}}, \bibinfo {author} {\bibfnamefont {F.}~\bibnamefont {Giazotto}}, \ and\ \bibinfo {author} {\bibfnamefont {G.~P.}\ \bibnamefont {Pepe}},\ }\href {\doibase 10.1103/PhysRevApplied.10.024015} {\bibfield  {journal} {\bibinfo  {journal} {Phys. Rev. Appl.}\ }\textbf {\bibinfo {volume} {10}},\ \bibinfo {pages} {024015} (\bibinfo {year} {2018})}\BibitemShut {NoStop}%
\bibitem [{\citenamefont {Tinkham}(2004)}]{tinkham2004}%
  \BibitemOpen
  \bibfield  {author} {\bibinfo {author} {\bibfnamefont {M.}~\bibnamefont {Tinkham}},\ }\href@noop {} {\emph {\bibinfo {title} {Introduction to {{Superconductivity}}}}}\ (\bibinfo  {publisher} {Courier Corporation},\ \bibinfo {year} {2004})\BibitemShut {NoStop}%
\bibitem [{\citenamefont {Subero}\ \emph {et~al.}(2023)\citenamefont {Subero}, \citenamefont {Maillet}, \citenamefont {Golubev}, \citenamefont {Thomas}, \citenamefont {Peltonen}, \citenamefont {Karimi}, \citenamefont {{Mar{\'i}n-Su{\'a}rez}}, \citenamefont {Yeyati}, \citenamefont {S{\'a}nchez}, \citenamefont {Park},\ and\ \citenamefont {Pekola}}]{subero2023}%
  \BibitemOpen
  \bibfield  {author} {\bibinfo {author} {\bibfnamefont {D.}~\bibnamefont {Subero}}, \bibinfo {author} {\bibfnamefont {O.}~\bibnamefont {Maillet}}, \bibinfo {author} {\bibfnamefont {D.~S.}\ \bibnamefont {Golubev}}, \bibinfo {author} {\bibfnamefont {G.}~\bibnamefont {Thomas}}, \bibinfo {author} {\bibfnamefont {J.~T.}\ \bibnamefont {Peltonen}}, \bibinfo {author} {\bibfnamefont {B.}~\bibnamefont {Karimi}}, \bibinfo {author} {\bibfnamefont {M.}~\bibnamefont {{Mar{\'i}n-Su{\'a}rez}}}, \bibinfo {author} {\bibfnamefont {A.~L.}\ \bibnamefont {Yeyati}}, \bibinfo {author} {\bibfnamefont {R.}~\bibnamefont {S{\'a}nchez}}, \bibinfo {author} {\bibfnamefont {S.}~\bibnamefont {Park}}, \ and\ \bibinfo {author} {\bibfnamefont {J.~P.}\ \bibnamefont {Pekola}},\ }\href {\doibase 10.1038/s41467-023-43668-3} {\bibfield  {journal} {\bibinfo  {journal} {Nat. Commun.}\ }\textbf {\bibinfo {volume} {14}},\ \bibinfo {pages} {7924} (\bibinfo {year} {2023})}\BibitemShut {NoStop}%
\bibitem [{\citenamefont {Gunnarsson}\ \emph {et~al.}(2015)\citenamefont {Gunnarsson}, \citenamefont {{Richardson-Bullock}}, \citenamefont {Prest}, \citenamefont {Nguyen}, \citenamefont {Timofeev}, \citenamefont {Shah}, \citenamefont {Whall}, \citenamefont {Parker}, \citenamefont {Leadley}, \citenamefont {Myronov},\ and\ \citenamefont {Prunnila}}]{gunnarsson2015}%
  \BibitemOpen
  \bibfield  {author} {\bibinfo {author} {\bibfnamefont {D.}~\bibnamefont {Gunnarsson}}, \bibinfo {author} {\bibfnamefont {J.~S.}\ \bibnamefont {{Richardson-Bullock}}}, \bibinfo {author} {\bibfnamefont {M.~J.}\ \bibnamefont {Prest}}, \bibinfo {author} {\bibfnamefont {H.~Q.}\ \bibnamefont {Nguyen}}, \bibinfo {author} {\bibfnamefont {A.~V.}\ \bibnamefont {Timofeev}}, \bibinfo {author} {\bibfnamefont {V.~A.}\ \bibnamefont {Shah}}, \bibinfo {author} {\bibfnamefont {T.~E.}\ \bibnamefont {Whall}}, \bibinfo {author} {\bibfnamefont {E.~H.~C.}\ \bibnamefont {Parker}}, \bibinfo {author} {\bibfnamefont {D.~R.}\ \bibnamefont {Leadley}}, \bibinfo {author} {\bibfnamefont {M.}~\bibnamefont {Myronov}}, \ and\ \bibinfo {author} {\bibfnamefont {M.}~\bibnamefont {Prunnila}},\ }\href {\doibase 10.1038/srep17398} {\bibfield  {journal} {\bibinfo  {journal} {Sci. Rep.}\ }\textbf {\bibinfo {volume} {5}},\ \bibinfo {pages} {17398} (\bibinfo {year} {2015})}\BibitemShut {NoStop}%
\bibitem [{\citenamefont {Sergeev}\ and\ \citenamefont {Mitin}(2000)}]{sergeev2000}%
  \BibitemOpen
  \bibfield  {author} {\bibinfo {author} {\bibfnamefont {A.}~\bibnamefont {Sergeev}}\ and\ \bibinfo {author} {\bibfnamefont {V.}~\bibnamefont {Mitin}},\ }\href {\doibase 10.1103/PhysRevB.61.6041} {\bibfield  {journal} {\bibinfo  {journal} {Phys. Rev. B}\ }\textbf {\bibinfo {volume} {61}},\ \bibinfo {pages} {6041} (\bibinfo {year} {2000})}\BibitemShut {NoStop}%
\bibitem [{\citenamefont {Meschke}\ \emph {et~al.}(2004)\citenamefont {Meschke}, \citenamefont {Pekola}, \citenamefont {Gay}, \citenamefont {Rapp},\ and\ \citenamefont {Godfrin}}]{meschke2004}%
  \BibitemOpen
  \bibfield  {author} {\bibinfo {author} {\bibfnamefont {M.}~\bibnamefont {Meschke}}, \bibinfo {author} {\bibfnamefont {J.~P.}\ \bibnamefont {Pekola}}, \bibinfo {author} {\bibfnamefont {F.}~\bibnamefont {Gay}}, \bibinfo {author} {\bibfnamefont {R.~E.}\ \bibnamefont {Rapp}}, \ and\ \bibinfo {author} {\bibfnamefont {H.}~\bibnamefont {Godfrin}},\ }\href {\doibase 10.1023/B:JOLT.0000016733.75220.5d} {\bibfield  {journal} {\bibinfo  {journal} {J. Low Temp. Phys.}\ }\textbf {\bibinfo {volume} {134}},\ \bibinfo {pages} {1119} (\bibinfo {year} {2004})}\BibitemShut {NoStop}%
\bibitem [{\citenamefont {{Jos{\'e} Mart{\'i}nez-P{\'e}rez}}\ and\ \citenamefont {Giazotto}(2014)}]{martinez-perez2014}%
  \BibitemOpen
  \bibfield  {author} {\bibinfo {author} {\bibfnamefont {M.}~\bibnamefont {{Jos{\'e} Mart{\'i}nez-P{\'e}rez}}}\ and\ \bibinfo {author} {\bibfnamefont {F.}~\bibnamefont {Giazotto}},\ }\href {\doibase 10.1038/ncomms4579} {\bibfield  {journal} {\bibinfo  {journal} {Nat. Commun.}\ }\textbf {\bibinfo {volume} {5}},\ \bibinfo {pages} {3579} (\bibinfo {year} {2014})}\BibitemShut {NoStop}%
\bibitem [{\citenamefont {{Mart{\'i}nez-P{\'e}rez}}\ \emph {et~al.}(2015)\citenamefont {{Mart{\'i}nez-P{\'e}rez}}, \citenamefont {Fornieri},\ and\ \citenamefont {Giazotto}}]{martinez-perez2015}%
  \BibitemOpen
  \bibfield  {author} {\bibinfo {author} {\bibfnamefont {M.~J.}\ \bibnamefont {{Mart{\'i}nez-P{\'e}rez}}}, \bibinfo {author} {\bibfnamefont {A.}~\bibnamefont {Fornieri}}, \ and\ \bibinfo {author} {\bibfnamefont {F.}~\bibnamefont {Giazotto}},\ }\href {\doibase 10.1038/nnano.2015.11} {\bibfield  {journal} {\bibinfo  {journal} {Nat. Nanotechnol.}\ }\textbf {\bibinfo {volume} {10}},\ \bibinfo {pages} {303} (\bibinfo {year} {2015})}\BibitemShut {NoStop}%
\bibitem [{\citenamefont {Taskinen}\ and\ \citenamefont {Maasilta}(2006)}]{taskinen2006}%
  \BibitemOpen
  \bibfield  {author} {\bibinfo {author} {\bibfnamefont {L.~J.}\ \bibnamefont {Taskinen}}\ and\ \bibinfo {author} {\bibfnamefont {I.~J.}\ \bibnamefont {Maasilta}},\ }\href {\doibase 10.1063/1.2357555} {\bibfield  {journal} {\bibinfo  {journal} {Appl. Phys. Lett.}\ }\textbf {\bibinfo {volume} {89}},\ \bibinfo {pages} {143511} (\bibinfo {year} {2006})}\BibitemShut {NoStop}%
\bibitem [{\citenamefont {Savin}\ \emph {et~al.}(2001)\citenamefont {Savin}, \citenamefont {Ahopelto}, \citenamefont {Kivinen}, \citenamefont {Manninen}, \citenamefont {Pekola},\ and\ \citenamefont {Prunnila}}]{savin2001}%
  \BibitemOpen
  \bibfield  {author} {\bibinfo {author} {\bibfnamefont {A.}~\bibnamefont {Savin}}, \bibinfo {author} {\bibfnamefont {J.}~\bibnamefont {Ahopelto}}, \bibinfo {author} {\bibfnamefont {P.}~\bibnamefont {Kivinen}}, \bibinfo {author} {\bibfnamefont {A.}~\bibnamefont {Manninen}}, \bibinfo {author} {\bibfnamefont {J.}~\bibnamefont {Pekola}}, \ and\ \bibinfo {author} {\bibfnamefont {M.}~\bibnamefont {Prunnila}},\ }\href@noop {} {\bibfield  {journal} {\bibinfo  {journal} {The XXXV annual conference of the Finnish physical society}\ } (\bibinfo {year} {2001})}\BibitemShut {NoStop}%
\bibitem [{\citenamefont {Roddaro}\ \emph {et~al.}(2011)\citenamefont {Roddaro}, \citenamefont {Pescaglini}, \citenamefont {Ercolani}, \citenamefont {Sorba}, \citenamefont {Giazotto},\ and\ \citenamefont {Beltram}}]{roddaro2011}%
  \BibitemOpen
  \bibfield  {author} {\bibinfo {author} {\bibfnamefont {S.}~\bibnamefont {Roddaro}}, \bibinfo {author} {\bibfnamefont {A.}~\bibnamefont {Pescaglini}}, \bibinfo {author} {\bibfnamefont {D.}~\bibnamefont {Ercolani}}, \bibinfo {author} {\bibfnamefont {L.}~\bibnamefont {Sorba}}, \bibinfo {author} {\bibfnamefont {F.}~\bibnamefont {Giazotto}}, \ and\ \bibinfo {author} {\bibfnamefont {F.}~\bibnamefont {Beltram}},\ }\href {\doibase 10.1007/s12274-010-0077-6} {\bibfield  {journal} {\bibinfo  {journal} {Nano Res.}\ }\textbf {\bibinfo {volume} {4}},\ \bibinfo {pages} {259} (\bibinfo {year} {2011})}\BibitemShut {NoStop}%
\bibitem [{\citenamefont {Degheidy}\ \emph {et~al.}(2022)\citenamefont {Degheidy}, \citenamefont {Abuali},\ and\ \citenamefont {Elkenany}}]{degheidy2022}%
  \BibitemOpen
  \bibfield  {author} {\bibinfo {author} {\bibfnamefont {A.~R.}\ \bibnamefont {Degheidy}}, \bibinfo {author} {\bibfnamefont {A.~M.}\ \bibnamefont {Abuali}}, \ and\ \bibinfo {author} {\bibfnamefont {E.~B.}\ \bibnamefont {Elkenany}},\ }\href {\doibase 10.1149/2162-8777/ac79cc} {\bibfield  {journal} {\bibinfo  {journal} {ECS J. Solid State Sci. Technol.}\ }\textbf {\bibinfo {volume} {11}},\ \bibinfo {pages} {063016} (\bibinfo {year} {2022})}\BibitemShut {NoStop}%
\bibitem [{\citenamefont {Giazotto}\ and\ \citenamefont {{Mart{\'i}nez-P{\'e}rez}}(2012)}]{giazotto2012}%
  \BibitemOpen
  \bibfield  {author} {\bibinfo {author} {\bibfnamefont {F.}~\bibnamefont {Giazotto}}\ and\ \bibinfo {author} {\bibfnamefont {M.~J.}\ \bibnamefont {{Mart{\'i}nez-P{\'e}rez}}},\ }\href {\doibase 10.1038/nature11702} {\bibfield  {journal} {\bibinfo  {journal} {Nature}\ }\textbf {\bibinfo {volume} {492}},\ \bibinfo {pages} {401} (\bibinfo {year} {2012})}\BibitemShut {NoStop}%
\bibitem [{\citenamefont {Meschke}\ \emph {et~al.}(2006)\citenamefont {Meschke}, \citenamefont {Guichard},\ and\ \citenamefont {Pekola}}]{meschke2006}%
  \BibitemOpen
  \bibfield  {author} {\bibinfo {author} {\bibfnamefont {M.}~\bibnamefont {Meschke}}, \bibinfo {author} {\bibfnamefont {W.}~\bibnamefont {Guichard}}, \ and\ \bibinfo {author} {\bibfnamefont {J.~P.}\ \bibnamefont {Pekola}},\ }\href {\doibase 10.1038/nature05276} {\bibfield  {journal} {\bibinfo  {journal} {Nature}\ }\textbf {\bibinfo {volume} {444}},\ \bibinfo {pages} {187} (\bibinfo {year} {2006})},\ \bibinfo {note} {comment: 5 pages, 3 figures},\ \Eprint {http://arxiv.org/abs/cond-mat/0605678} {arxiv:cond-mat/0605678} \BibitemShut {NoStop}%
\bibitem [{\citenamefont {Buonomo}\ \emph {et~al.}(2003)\citenamefont {Buonomo}, \citenamefont {Leoni}, \citenamefont {Castellano}, \citenamefont {Mattioli}, \citenamefont {Torrioli}, \citenamefont {Di~Gaspare},\ and\ \citenamefont {Evangelisti}}]{buonomoE2003}%
  \BibitemOpen
  \bibfield  {author} {\bibinfo {author} {\bibfnamefont {B.}~\bibnamefont {Buonomo}}, \bibinfo {author} {\bibfnamefont {R.}~\bibnamefont {Leoni}}, \bibinfo {author} {\bibfnamefont {M.~G.}\ \bibnamefont {Castellano}}, \bibinfo {author} {\bibfnamefont {F.}~\bibnamefont {Mattioli}}, \bibinfo {author} {\bibfnamefont {G.}~\bibnamefont {Torrioli}}, \bibinfo {author} {\bibfnamefont {L.}~\bibnamefont {Di~Gaspare}}, \ and\ \bibinfo {author} {\bibfnamefont {F.}~\bibnamefont {Evangelisti}},\ }\href {\doibase 10.1063/1.1627952} {\bibfield  {journal} {\bibinfo  {journal} {J. Appl. Phys.}\ }\textbf {\bibinfo {volume} {94}},\ \bibinfo {pages} {7784} (\bibinfo {year} {2003})}\BibitemShut {NoStop}%
\bibitem [{\citenamefont {Giazotto}\ \emph {et~al.}(2008)\citenamefont {Giazotto}, \citenamefont {Heikkil{\"a}}, \citenamefont {Pepe}, \citenamefont {Helist{\"o}}, \citenamefont {Luukanen},\ and\ \citenamefont {Pekola}}]{giazotto2008}%
  \BibitemOpen
  \bibfield  {author} {\bibinfo {author} {\bibfnamefont {F.}~\bibnamefont {Giazotto}}, \bibinfo {author} {\bibfnamefont {T.~T.}\ \bibnamefont {Heikkil{\"a}}}, \bibinfo {author} {\bibfnamefont {G.~P.}\ \bibnamefont {Pepe}}, \bibinfo {author} {\bibfnamefont {P.}~\bibnamefont {Helist{\"o}}}, \bibinfo {author} {\bibfnamefont {A.}~\bibnamefont {Luukanen}}, \ and\ \bibinfo {author} {\bibfnamefont {J.~P.}\ \bibnamefont {Pekola}},\ }\href {\doibase 10.1063/1.2908922} {\bibfield  {journal} {\bibinfo  {journal} {Appl. Phys. Lett.}\ }\textbf {\bibinfo {volume} {92}},\ \bibinfo {pages} {162507} (\bibinfo {year} {2008})}\BibitemShut {NoStop}%
\end{thebibliography}%

\end{document}